\documentstyle[twoside,epsf]{article}

\catcode`\@=11
\long\def\@makefntext#1{
\protect\noindent \hbox to 3.2pt {\hskip-.9pt
$^{{\eightrm\@thefnmark}}$\hfil}#1\hfill}

\def\@makefnmark{\hbox to 0pt{$^{\@thefnmark}$\hss}}
\def\ps@myheadings{\let\@mkboth\@gobbletwo
\def\@oddhead{\hbox{}
\rightmark\hfil\eightrm\thepage}
\def\@oddfoot{}\def\@evenhead{\eightrm\thepage\hfil
\leftmark\hbox{}}\def\@evenfoot{}
\def\sectionmark##1{}\def\subsectionmark##1{}}

\oddsidemargin=\evensidemargin
\newcounter{sectionc}\newcounter{subsectionc}\newcounter{subsubsectionc}
\renewcommand{\section}[1] {\vspace{12pt}\addtocounter{sectionc}{1}
\setcounter{subsectionc}{0}\setcounter{subsubsectionc}{0}\noindent
    {\tenbf\thesectionc. #1}\par\vspace{5pt}}
\renewcommand{\subsection}[1] {\vspace{12pt}\addtocounter{subsectionc}{1}
\setcounter{subsubsectionc}{0}\noindent
{\bf\thesectionc.\thesubsectionc. {\kern1pt \bfit #1}}\par\vspace{5pt}}
\renewcommand{\subsubsection}[1] {\vspace{12pt}\addtocounter{subsubsectionc}{1}
    \noindent{\tenrm\thesectionc.\thesubsectionc.\thesubsubsectionc.
    {\kern1pt \tenit #1}}\par\vspace{5pt}}
\newcommand{\nonumsection}[1] {\vspace{12pt}\noindent{\tenbf #1}
    \par\vspace{5pt}}
\newcounter{appendixc}
\newcounter{subappendixc}[appendixc]
\newcounter{subsubappendixc}[subappendixc]
\renewcommand{\thesubappendixc}{\Alph{appendixc}.\arabic{subappendixc}}
\renewcommand{\thesubsubappendixc}
    {\Alph{appendixc}.\arabic{subappendixc}.\arabic{subsubappendixc}}
\renewcommand{\appendix}[1] {\vspace{12pt}
        \refstepcounter{appendixc}
        \setcounter{figure}{0}
        \setcounter{table}{0}
        \setcounter{lemma}{0}
        \setcounter{theorem}{0}
        \setcounter{corollary}{0}
        \setcounter{definition}{0}
        \setcounter{equation}{0}
        \renewcommand{\thefigure}{\Alph{appendixc}.\arabic{figure}}
        \renewcommand{\thetable}{\Alph{appendixc}.\arabic{table}}
        \renewcommand{\theappendixc}{\Alph{appendixc}}
        \renewcommand{\thelemma}{\Alph{appendixc}.\arabic{lemma}}
        \renewcommand{\thetheorem}{\Alph{appendixc}.\arabic{theorem}}
        \renewcommand{\thedefinition}{\Alph{appendixc}.\arabic{definition}}
        \renewcommand{\thecorollary}{\Alph{appendixc}.\arabic{corollary}}
        \renewcommand{\theequation}{\Alph{appendixc}.\arabic{equation}}
        \noindent{\tenbf Appendix \theappendixc #1}\par\vspace{5pt}}
\newcommand{\subappendix}[1] {\vspace{12pt}
        \refstepcounter{subappendixc}
        \noindent{\bf Appendix \thesubappendixc. {\kern1pt \bfit #1}}
    \par\vspace{5pt}}
\newcommand{\subsubappendix}[1] {\vspace{12pt}
        \refstepcounter{subsubappendixc}
        \noindent{\rm Appendix \thesubsubappendixc. {\kern1pt \tenit #1}}
    \par\vspace{5pt}}
\newcommand{\textlineskip}{\baselineskip=13pt}
\newcommand{\smalllineskip}{\baselineskip=10pt}

\newcommand{\copyrightheading}[1]
    {\vspace*{-2.5cm}\smalllineskip{\flushleft
    {\footnotesize Quantum Information and Computation, Vol.~0, No.~0 (2001) 000--000 #1}\\
    {\footnotesize \copyright\kern2pt Rinton Press}\\
     }}

\def\abstracts#1#2#3{{
    \centering{\begin{minipage}{4.5in}\footnotesize\baselineskip=10pt
    \parindent=0pt #1\par
    \parindent=15pt #2\par
    \parindent=15pt #3
    \end{minipage}}\par}}

\def\keywords#1{{
    \centering{\begin{minipage}{4.5in}\footnotesize\baselineskip=10pt
    {\footnotesize\it Keywords}\/: #1
     \end{minipage}}\par}}

\renewenvironment{thebibliography}[1]
        {\frenchspacing
     \ninerm\baselineskip=11pt
         \begin{list}{\arabic{enumi}.}
        {\usecounter{enumi}\setlength{\parsep}{0pt}
     \setlength{\leftmargin 12.7pt}{\rightmargin 0pt}
         \setlength{\itemsep}{0pt} \settowidth
    {\labelwidth}{#1.}\sloppy}}{\end{list}}

\newcounter{itemlistc}
\newcounter{romanlistc}
\newcounter{alphlistc}
\newcounter{arabiclistc}

\newcommand{\fcaption}[1]{
        \refstepcounter{figure}
        \setbox\@tempboxa = \hbox{\footnotesize Fig.~\thefigure. #1}
        \ifdim \wd\@tempboxa > 5in
           {\begin{center}
        \parbox{5in}{\footnotesize\smalllineskip Fig.~\thefigure. #1}
            \end{center}}
        \else
             {\begin{center}
             {\footnotesize Fig.~\thefigure. #1}
              \end{center}}
        \fi}

\newcommand{\tcaption}[1]{
        \refstepcounter{table}
        \setbox\@tempboxa = \hbox{\footnotesize Table~\thetable. #1}
        \ifdim \wd\@tempboxa > 5in
           {\begin{center}
        \parbox{5in}{\footnotesize\smalllineskip Table~\thetable. #1}
            \end{center}}
        \else
             {\begin{center}
             {\footnotesize Table~\thetable. #1}
              \end{center}}
        \fi}

\def\pmb#1{\setbox0=\hbox{#1}
    \kern-.025em\copy0\kern-\wd0
    \kern.05em\copy0\kern-\wd0
    \kern-.025em\raise.0433em\box0}

\def\fnt#1#2{\footnotetext{\kern-.3em
    {$^{\mbox{\scriptsize #1}}$}{#2}}}

\def\fpage#1{\begingroup
\voffset=.3in
\thispagestyle{empty}\begin{table}[b]\centerline{\footnotesize #1}
    \end{table}\endgroup}

\def\runninghead#1#2{\pagestyle{myheadings}
\markboth{{\protect\footnotesize\it{\quad #1}}\hfill}
{\hfill{\protect\footnotesize\it{#2\quad}}}}
\font\tenrm=cmr10
\font\tenit=cmti10
\font\tenbf=cmbx10
\font\bfit=cmbxti10 at 10pt
\font\ninerm=cmr9

\font\eightrm=cmr8

\def\FigName{figure}%
\newbox\captionbox
\long\def\@makecaption#1#2{%
  \ifx\FigName\@captype
    \vskip\abovecaptionskip
    \setbox\tempbox\hbox{{\figurecaptionfont #1\hskip1em #2}}
    \ifdim\wd\tempbox< 28pc
    \centerline{\box\tempbox}
    \else
    {\figurecaptionfont #1\hskip1em #2\par}
\fi\else
    \setbox\tempbox\hbox{{\tablecaptionfont #1\hskip1em #2}}
    \ifdim\wd\tempbox< 28pc
    \centerline{\box\tempbox}
    \else
    {\tablecaptionfont #1\hskip1em #2\par}%
    \fi
 \vskip\belowcaptionskip
 \fi}
\InputIfFileExists{psfig.sty}
{\typeout{^^Jpsfig.sty inputed...ok}}{\typeout{^^JWarning: psfig.sty could be be found.^^J}}
\InputIfFileExists{epsfsafe.tex}
{\typeout{^^Jepsfsafe.tex inputed...ok}}
            {\typeout{^^JWarning: epsfsafe.tex could not be found.^^J}}
\InputIfFileExists{epsfig.sty}
{\typeout{^^Jepsfig.sty inputed...ok}}{\typeout{^^JWarning: epsfig.sty could not be found.^^J}}
\InputIfFileExists{epsf.sty}
{\typeout{^^Jepsf.sty inputed...ok}}{\typeout{^^JWarning: epsf.sty could not be found.^^J}}%
\def\fps@figure{tbp}
\def\ftype@figure{1}
\def\ext@figure{lof}
\def\fnum@figure{Fig.\ \thefigure}
\def\qed{\hbox{${\vcenter{\vbox{              
   \hrule height 0.4pt\hbox{\vrule width 0.4pt height 6pt
   \kern5pt\vrule width 0.4pt}\hrule height 0.4pt}}}$}}

\def\<{\langle}
\def\>{\rangle}
\def\be{\begin{equation}}
\def\ee{\end{equation}}
\def\bea{\begin{eqnarray}}
\def\eea{\end{eqnarray}}

\begin{document}

\setlength{\textheight}{8.0truein}    

\runninghead{Experimental realization of entangled qutrits for
quantum communication}
            {R.~Thew, A.~Ac\'in, H.~Zbinden and N.~Gisin}

\normalsize\textlineskip
\thispagestyle{empty}
\setcounter{page}{1}


\vspace*{0.88truein}

\fpage{1}


\centerline{\bf EXPERIMENTAL REALIZATION OF }
\vspace*{0.015truein} \centerline{\bf ENTANGLED QUTRITS FOR
QUANTUM COMMUNICATION} \vspace*{0.37truein}


\centerline{\footnotesize R.~Thew$^1$ A.~Ac\'in$^{1,2}$,
H.~Zbinden$^1$ and N.~Gisin$^1$} \vspace*{0.15truein}
\centerline{\footnotesize\it $^1$Group of Applied Physics,
University of Geneva, 1211 Geneva 4, Switzerland}
 \centerline{\footnotesize\it and } \baselineskip=10pt
\centerline{\footnotesize\it $^2$Institut de Ci\`encies
Fot\`oniques, Jordi Girona 29, 08034 Barcelona, Spain}
 \vspace*{0.225truein}
\vspace*{10pt}


 \vspace*{10pt}

\vspace*{0.21truein}
%

\abstracts{We have experimentally realized a technique to
generate, control and measure entangled qutrits, 3-dimensional
quantum systems. This scheme uses spontaneous parametric down
converted photons and unbalanced 3-arm fiber optic interferometers
in a scheme analogous to the Franson interferometric arrangement
for qubits. The results reveal a source capable of generating
maximally entangled states with a net state fidelity, F = 0.985
$\pm$ 0.018. Further the control over the system reveals a high,
net, 2-photon interference fringe visibility, V = 0.919 $\pm$
0.026. This has all been done at
telecom wavelengths thus facilitating the advancement towards long
distance higher dimensional quantum communication.} {}{}

\vspace*{10pt}

\vspace*{3pt}




\setcounter{footnote}{0}

\vspace*{1pt}\textlineskip


\section{Introduction}

Quantum information science is now a well recognized branch of
science in its own right, dealing with a diverse range of
theoretical and practical issues related to quantum computing and
quantum communication \cite{Nielsen00a}. At the heart of this
field has been the idea of the Qubit, simply, a two dimensional
quantum system. Theory and experiment, especially within the
optics community, has advanced rapidly, both from a fundamental as
well as applied perspective. Already we have quantum key
distribution (QKD) \cite{Gisin01a} as the first quantum technology
to be taken into the public domain. One question that has received
significant theoretical interest in the past few years is - what
can we do with high dimensional systems?

There have recently been several proposals for quantum
communication protocols involving higher dimensional states,
specifically qutrits (three dimensional quantum systems). In some
cases these have advantages over qubit protocols, like increased
robustness against noise in quantum key distribution schemes
\cite{Bechmann00a,Cerf02a,Brub02a}. However, the more interesting
cases are those where it is more efficient to have a system
encoded using qutrits than qubits, specifically for the solution
to the Byzantine Agreement Problem \cite{Fitzi01a} or for Quantum
Coin Tossing \cite{Ambainis02a}. As such we would like an
efficient source of entangled qutrits to take advantage of these
protocols. Also, if we can approach these qutrit and higher
dimensional systems with the same positive perspective that has
been accorded the qubit then what we have are larger quantum
system in which to perform even more complex quantum protocols.

There are two possible approaches that can be taken if we want to
investigate higher dimensional systems: use multiple  (more than
2)  entangled qubit systems
\cite{Sacket00a,Pan00a,Jennewein02a,Howell02a,Marcikic03a}; or
increase the dimensions of the fundamental elements. If we
increase the dimensions of the elements, the simplest place to
start to answer some of the questions experimentally is with
qutrits. Recently there have been some advances to this end: the
generation of qutrits using bi-photons \cite{Trifonov00a,Burlakov02a}; and two
schemes generating entangled qudits, $d$-level quantum systems.
The first, using the orbital angular momentum of light, have
specifically shown their scheme setup to analyze qutrits
\cite{Vaziri02a}, while the second uses a pulsed, mode-locked,
laser to generate time-bin qudits where they have shown
entanglement up to $d = 11$ \cite{deRiedmatten02a}.

The scheme we use is based on energy-time entanglement and is
analogous to the Franson-type interferometric arrangement for
photonic qubits \cite{Franson89a}. A similar idea for such an
experiment had previously been proposed however only preliminary
experimental results followed \cite{Reck96a,Weihs96a,Weihs96b}. In
this
 article we introduce an experimental setup where we can
generate, manipulate and measure entangled qutrits. We first
detail our experimental set-up and show how this corresponds to
the theoretical description of qutrits before deriving two
different means of analyzing the output of the system.

\section{Experimental Set-up}

\begin{figure}\leavevmode
\begin{center}
\epsfig{figure=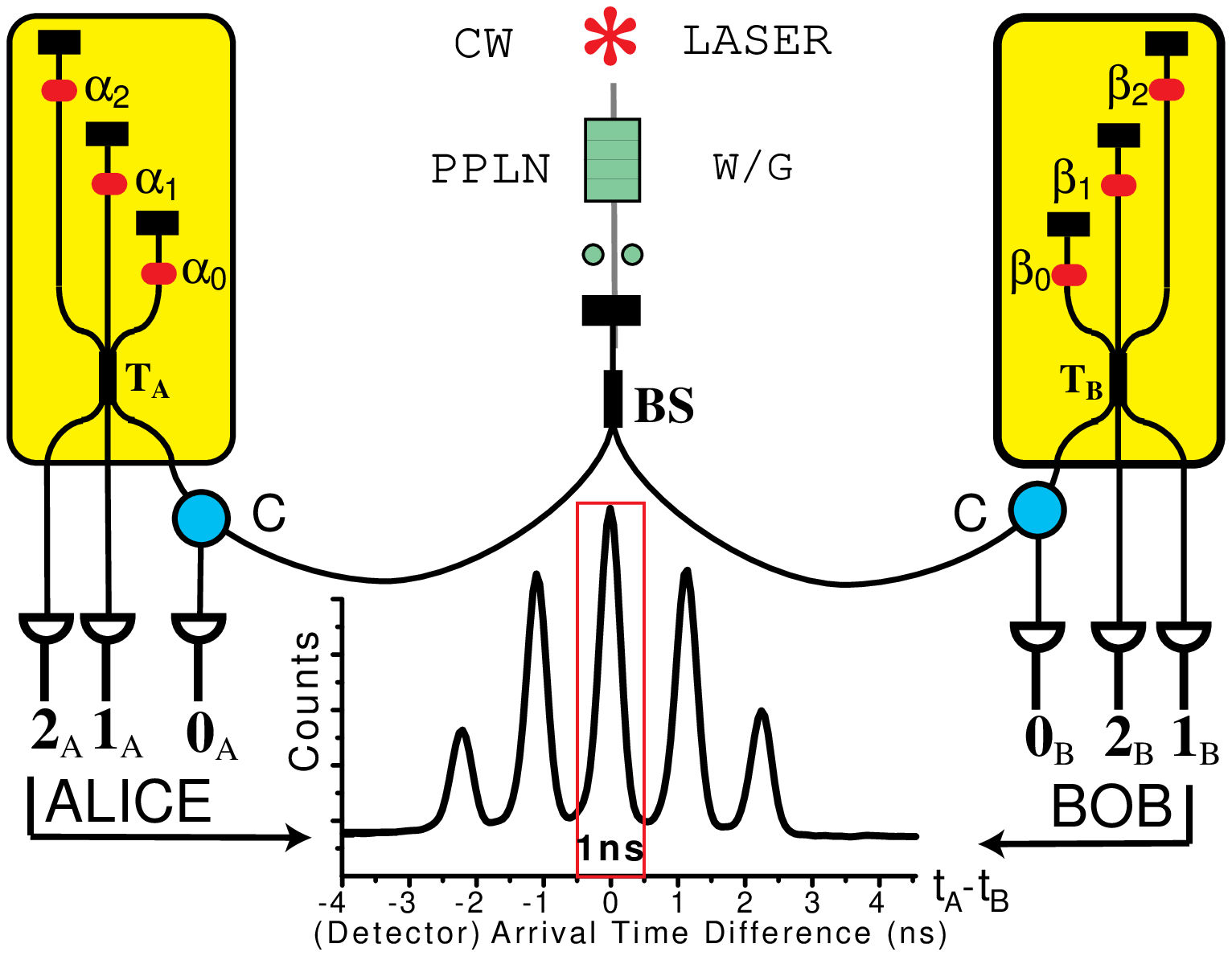,width=90mm}
\end{center}
\fcaption{\label{fig:qutritfig1} Schematic of experiment for
generation and analysis of entangled qutrits: Down-Converted
photon pairs at telecom wavelengths combined with unbalanced three
path interferometers are used to generate and analyze energy-time
entangled qutrits. See text for details.}
\end{figure}

Consider the experimental schematic of Fig.\ref{fig:qutritfig1}.
We are using a continuous wave (CW) external grating  diode laser
(Toptica DL100) at 657\,nm incident on a Periodically Poled
Lithium Niobate (PPLN) waveguide (Uni. of Nice) where spontaneous Parametric Down
Conversion (PDC) produces degenerate, collinear,  energy-time
entangled photon pairs at 1314\,nm.  The PPLN waveguide has proven
to be a highly efficient and stable source of PDC photons
\cite{Tanzilli01a} which allows one to use a simple laser diode
instead of  a large, complex, and expensive laser system often
required when using standard non-linear crystals. The photon pairs
are coupled into a monomode optical fiber before passing through a
fiber beam splitter (BS) which separates the photon pairs and
sends them to the two all-fiber Michelson interferometers.

Each of the interferometers consist of a 6-port symmetric coupler
\cite{Zukowski97a} T$_{A}$ and T$_{B}$, also known as tritters.
These are the three dimensional extension to a standard beam
splitter or 50/50 fiber coupler. Ideally they take an input signal
in any of the three inputs and distribute it with equal
probability in the three output ports and hence we would like
a 33/33/33 fiber coupler. If we can satisfy this condition we can make the assumption that our source generates symmetric, maximally entangled, qutrits \cite{Zukowski97a}. The splitting ratios have been measured
for both of these and we find that the deviation from this ideal
value is around 5\%. Even a variation of 10\%  would only
correspond to a reduction in state fidelity of less than 1\%. The
fidelity is not the most sensitive measure \cite{Peters03a} but
this symmetry is crucial in simplifying the analysis of the system
as we will see momentarily.

When the photons enter the interferometers via the tritters
T$_{A}$ and T$_{B}$, they have a choice of three paths. The three
possibilities hence define the qutrit space. We encode these
possible paths via the path-lengths, ie. short, medium, long, or
0,1,2, as labelled. We define the three different path lengths
such that the path-length differences in both interferometers are
the same, within the coherence length of the PDC photons. More
formally we require that the path lengths satisfy,
\begin{eqnarray}
\begin{array}{ccccccc}
 l^{A} - m^{A} & \approx  & m^{A} - s^{A}  &  \approx &  l^{B} - m^{B}
& \approx & m^{B} - s^{B}
\end{array}
\end{eqnarray}
where, for example,  $m^{A}$ denotes the length of fibre for the
medium arm in Alice's interferometer. The path length differences
define the separation of the five peaks in the time-of-arrival
histogram inset into Fig.\ref{fig:qutritfig1}, which has been
chosen to be $\Delta \tau = 1.2\,ns$.  Although we are encoding
our qutrits with time-bins we also need to be careful with the
polarization state of the photons. Inside the fiber interferometer
the polarization can vary and hence one could possibly determine
which path was taken by the photon. To overcome this problem we
use Faraday Mirrors at the ends of these fiber arms so that when a
photon returns to the tritter its polarization is always
orthogonal to when it entered, {\it regardless of the path taken},
thus reducing the which-path information. Any which-path
information could in turn diminish the quality of the entanglement
resource.

When the photons leave the interferometers there are nine possible
detection combinations that can be observed between each of Alice
and Bob's three detectors. We can put optical Circulators (C) on
one input port of each interferometer so we can detect all of
these. As well as this there are nine different path combinations
through the two interferometers for each detection combination.
These are represented in the time-of-arrival histogram inset in
Fig.\ref{fig:qutritfig1}. The two outside peaks correspond to a
difference in the arrival time of the photon pairs of $\pm 2\Delta
\tau$, where one photon passes through a short arm and the other
pases through a long arm. The next two lateral peaks, either side
of the central peak, correspond to photons passing through the
short and medium {\it or} the medium and long arms and arriving
with $\pm \Delta \tau$ and thus being on the left or right of the
central peak. Finally, the central peak corresponds to where the
photons take the same path in each interferometer, short-short,
medium-medium, or long-long and arrive with no time difference.

We can only discuss these events in this manner under the
following circumstances: the coherence length of the PDC photons
is much smaller than the path-length difference in the
interferometers (45\,$\mu$m $<<$ 24\,cm) so that no single photon
interference effects are observed in passing the interferometers;
the coherence length of the pump laser  is much greater than these
path-length differences (48\,cm $<<$ $\cal{O}$(100\,m) ) so that
we have no timing information as to the creation time of the
photon pair and hence which path was taken before detection.

 If we do this we can then describe the experiment as follows. We use energy-time correlated photon pairs distributed to two tritters as our source of qutrits. These qutrits are maximally entangled due to the symmetry of the coupling ratio of the tritters, as previously discussed. The different path-lengths in the interferometers defines a transform basis and provides a means of individually addressing the two free phases with local phase operations. Equivalently, the post-selection defines a well defined reference time, say relative to Alice. This, a priori, defines a superposition of three coherent creation times for photons that will arrive in coincidence after passing the short, medium and long paths of  the interferometers. We can vary two independent phases in each interferometer by changing their relative path lengths. The photons leaving the interferometers via their respective tritters
and then being detected can be viewed as a projective measurement onto the entangled qutrit state space. The result depending on the interference effects from the three indistinguishable  paths  and the phases applied.

Therefore, with these assumptions we can define three orthogonal,
maximally entangled, qutrit states for the three different possible
coincidence detection combinations.
\begin{eqnarray}
|\psi_{00}\rangle &=&\frac{1}{\sqrt{3}}[ |00\rangle +
e^{i(\alpha_{1}-\alpha_{0}+\beta_{1}-\beta_{0}+t)}|11\rangle
 \nonumber \\ && \hspace{2.5cm} +
 e^{i(\alpha_{2}-\alpha_{0}+\beta_{2}-\beta_{0}+t)}|22 \rangle ] \label{eq:eq1} \\
|\psi_{01}\rangle &=&
\frac{1}{\sqrt{3}}[|00\rangle+e^{i(\alpha_{1}-\alpha_{0}+\beta_{1}-\beta_{0})}|11\rangle
\nonumber \\ && \hspace{2.5cm}
+ e^{i(\alpha_{2}-\alpha_{0}+\beta_{2}-\beta_{0}-t)}|22 \rangle ] \\
|\psi_{02}\rangle &=& \frac{1}{\sqrt{3}}[|00\rangle +
e^{i(\alpha_{1}-\alpha_{0}+\beta_{1}-\beta_{0}-t)}|11\rangle
\nonumber \\ && \hspace{2.5cm}+
e^{i(\alpha_{2}-\alpha_{0}+\beta_{2}-\beta_{0})}|22 \rangle ].
\end{eqnarray}
Here $|\psi_{01}\rangle $ describes the state, determined by the
interferometer paths and phase settings, that leads to a
coincidence detection between Alice's detector 0 and Bob's
detector 1. $ |11\rangle= |1\rangle_{A} \otimes |1\rangle_{B}$
corresponds to the photon taking the medium path in both
interferometers. The phases that can be applied in each arm of the
interferometers are denoted by $\alpha_{j}, \beta_{j}$. We have
also introduced the factor $t = 2 \pi /3$ which is the phase
obtained when a photon changes ports in the tritters, analogous to
the $\pi/2$ phase obtained upon reflection in normal 4-port
beam-splitters. The previously mentioned symmetry ensures that the
states satisfy, $|\psi_{00}\rangle= |\psi_{11}\rangle =
|\psi_{22}\rangle$, $|\psi_{01}\rangle = |\psi_{12}\rangle =
|\psi_{20}\rangle$, and $|\psi_{02}\rangle = |\psi_{10}\rangle =
|\psi_{21}\rangle $.

Qutrit QKD, as mentioned in the introduction, can provide a more
robust key distribution scheme than with qubits. It also provides
another means to formulate a description of this experimental
arrangement that may be more intuitive. Alice prepares her qutrit
in one of the three transpose bases with her interferometer,
generating a superposition of the three (computational) basis
states, and her choice of phase settings \cite{Bechmann00a}. She
will obtain a click at one of three detectors, her trit value. She
sends the other,''prepared'' photon to Bob. Bob makes a choice
about which basis with his interferometer and phase settings
defining one of his three transform basis choices. He receives a
click at one of his three detectors, his trit value. If their
bases agree then the results are correlated and they can share a
secret key. We are performing this experiment at 1.3\,$\mu$m and
this type of encoding has already been shown to provide a robust
form of entanglement  for long distance quantum communication
\cite{Tittel99a,Thew02a} hence there are very few restrictions in
extending the current scheme to longer distances for protocols of
this type.

For detection we are using a combination of Ge and InGaAs
Avalanche Photo-Diodes (APDs). The Ge work in a passive mode at
1314\,nm and have efficiencies, $\eta_{Ge}\approx 10 \% $. We use
these on one side, Alice's, to trigger the InGaAs detectors
(idQuantique id200) on Bob's side which need to be used in a gated
mode but have a higher quantum efficiency, $\eta_{InGaAs}
> 20 \% $, and better noise characteristics. We use a
Time-To-Digital converter (ACAM) to process these detection events which
generates the data corresponding to arrival time differences between
the start (Alice) and the three different stops (Bob). For each of
these start-stop combinations we obtain an histogram like the one
inset between Alice and Bob's detectors in
Fig.\ref{fig:qutritfig1}. We select events occurring
within temporal windows $\Delta \tau_{w}$ = 1\,ns about these
peaks. Events in each central peak of the three histograms corresponds to a
projection onto  one of the three states previously listed.

The other peaks here are not without interest. The first two
lateral peaks project onto different subspaces within the
entangled qutrit Hilbert space. Specifically, detecting events in
these peaks projects onto states of the form,
\begin{eqnarray}
|\psi_{00}^R\rangle&=&  \frac{1}{\sqrt{2}}[|01\rangle +
e^{i(\alpha_{1}-\alpha_{0}+\beta_{2}-\beta_{1}-t)}|12\rangle  ]
\end{eqnarray}
for the right peak where the photons arrive with a fixed time
difference $\Delta \tau = 1.2$\,ns. There are two
possibilities and we cannot distinguish photons that took the
short(0) path at Alice's and the medium(1) path at Bob's from
those that took the medium(1) at Alice's and the long(2) at Bob's.
The left peak corresponds to those states having an arrival time
difference of $-\Delta \tau$, and these project onto states of the
form
\begin{eqnarray}
|\psi_{00}^L\rangle&=& \frac{1}{\sqrt{2}}[|10\rangle +
e^{i(\alpha_{2}-\alpha_{1}+\beta_{1}-\beta_{0}-t)}|21\rangle  ].
\end{eqnarray}
Both left and right states here are entangled and their form is
suggestive of the type of states required for optimal quantum coin
tossing \cite{Ambainis02a}. We have the same symmetry for the
other detectors, as previously shown for the states in the central
peak, and again these results differ by factors of $t$. The outer
two peaks, $\pm 2 \Delta \tau$ difference in arrival times, ie.
$|02\rangle$ and $|20\rangle$, have no interfering effects and
hence can be used to monitor  the count rate. 
Finally the noise consists of accidental coincidences due to detector noise and
uncorrelated photons and is measured concurrently with the true
coincidence counts,  by looking at the events which occur outside of these five peaks.

We commented that states corresponding to these lateral peaks were
interesting but they are also useful in controlling and
characterizing the entangled qutrits of the central peak. We will
vary the phases in just one, Alice's, interferometer, thus, to
simplify the theory  we can
think of Bob's interferometer as a reference and set all of these
phase settings to zero, $(\beta_{0}, \beta_{1}, \beta_{2})=
(0,0,0)$.  If we do this we find that the lateral peaks now have
corresponding states of the form:
\begin{eqnarray}
|\psi_{00}^R\rangle &=& |01\rangle + e^{i(\Phi_{00}^R+t)}|12\rangle
\hspace{1cm} : \hspace{1cm}   \Phi_{00}^R \equiv \alpha_{1}-\alpha_{0} + t \nonumber  \\
    |\psi_{00}^L\rangle &=&  |10\rangle + e^{i(\Phi_{00}^L- t)}|21\rangle
    \hspace{1cm} : \hspace{1cm} \Phi_{00}^L \equiv \alpha_{2}-\alpha_{1}   \label{eqLRphasestates},
\end{eqnarray}
and similarly for the other two. We can then rewrite
the entangled qutrit state of Eq.(\ref{eq:eq1}) as
\begin{eqnarray}
|\psi_{00}\rangle &= &\frac{1}{\sqrt{3}}[|00\rangle + e^{i
\Phi_{00}^R}|11\rangle +  e^{i(\Phi_{00}^R+\Phi_{00}^L)}|22
\rangle ].
\end{eqnarray}
In fact all three different states can be written in this form
with the phase shift of $t$ within the definition of the phase of
the lateral peaks. Thus, if we are trying to align the
interferometers or determine some relative phases we can observe
the various probabilities of coincidence in the associated peak of
the time-of-arrival histogram to assist us. By observing the
behavior of these lateral peaks we can control the relationship
between the two phases.

In the interferometers we control the phase, its setting and
stability, via temperature. As the fiber is heated it gets longer
and thus the phase relative to the other fiber arms is shifted.
For instance when we are scanning the temperature/phase in one
interferometer we have a temperature dependence of around
$9\pi/K$. Both interferometers are actively temperature stabilized
and the interferometer that is left fixed is clearly stable over
several periods as can be seen by the repeatability of the
interference fringes inset in Fig.\ref{fig:visibilities}. By
looking at the interference fringes in the lateral peaks we can
determine the relationship between the two free phases and hence
we can vary them with a fixed, and known,  relationship, e.g.
$\Phi_{jk}^R= r\Phi_{jk}^L$,  for some constant coefficient, $r$.

\section{Characterization}

So how can we characterize and quantify these states? From the
perspective of the optician we would like to be able to define
some interference fringe visibility to quantify the process. This
can always be defined as $ V =(I_{max}-I_{min})/(I_{max}+I_{min})$
where, in the case of single photon detection, we can replace
these maximum and minimum intensities with detection
probabilities, with the limits taken over all possible phases.
Before we determine the detection probabilities for these qutrit
states we have defined  we will  introduce a standard, symmetric,
noise model such that we can consider the case where we have,
\begin{eqnarray}
    \rho_{jk} = \lambda|\psi_{jk}\rangle \langle \psi_{jk}| +
    (1-\lambda)I_{9}/9\label{eq:mixed}.
\end{eqnarray}
With probability $\lambda$ we expect our entangled qutrit state,
and with probability $(1-\lambda)$ we expect a maximally mixed
state, at output $jk$. A symmetric model is assumed at the outset
for its simplicity and due to the high degree of symmetry for the
tritter coupling ratios. As we will see this theory fits very well
with the experimental results and hence is justified.

\begin{figure}\leavevmode
\begin{center}
\epsfig{figure=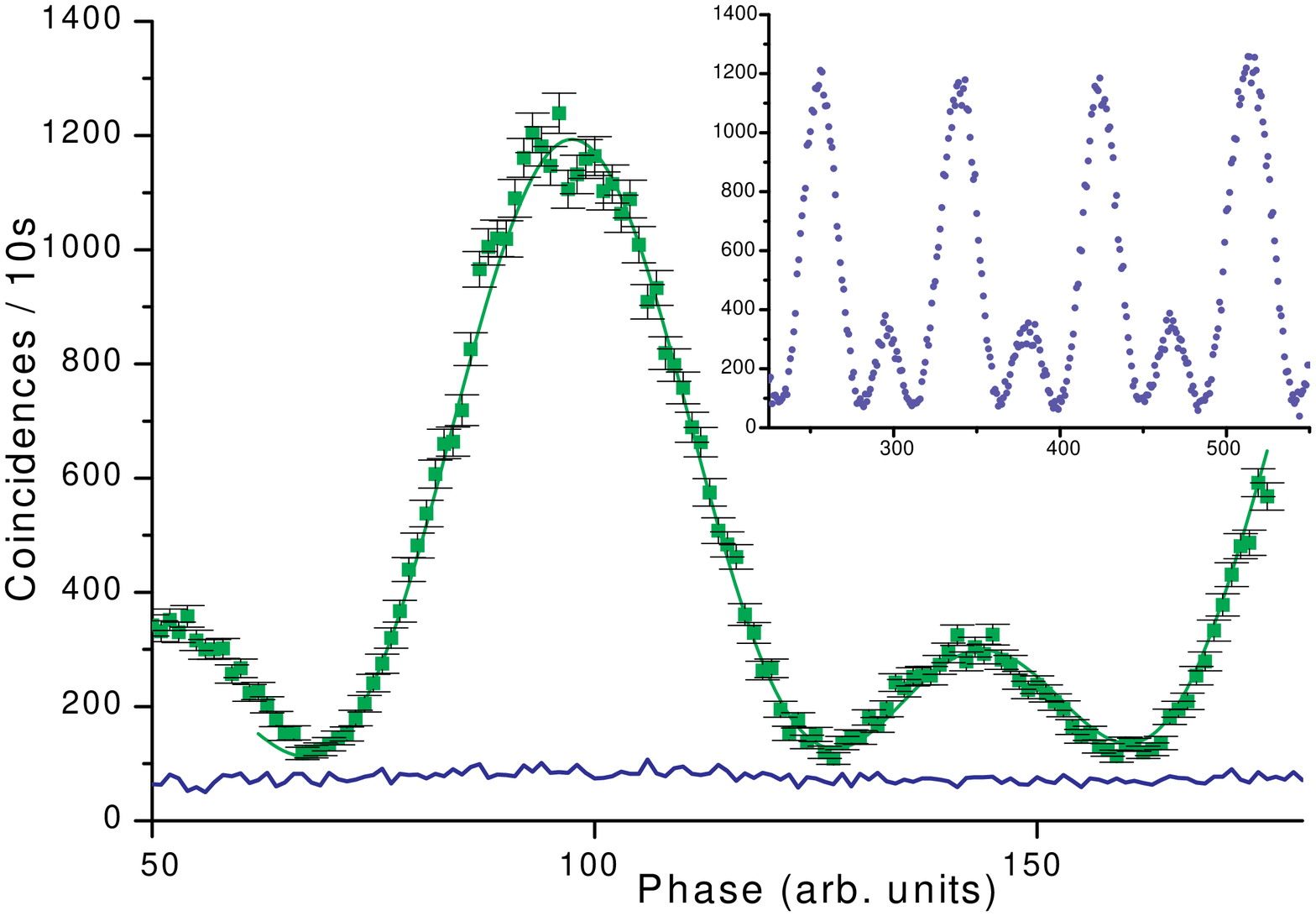,width=100mm}
\end{center}
\fcaption{\label{fig:visibilities}Raw interference fringe,
coincidence counts, for the central peak. The accidental
coincidence (noise) level is also shown. Inset we see another
interference pattern showing the stability and repeatability of
the source.}
\end{figure}
If we do this we have the following probability for a coincidence
detection projecting onto the mixed state of Eq.(\ref{eq:mixed}),
\begin{eqnarray}
   P_{jk} = \frac{1}{3}\left[ 3 + 2\lambda \left[\cos(\Phi_{jk}^R) + \cos(\Phi_{jk}^L) + \cos(\Phi_{jk}^R+\Phi_{jk}^L)\right]\right].\label{eq:fringes}
\end{eqnarray}
We can clearly see the dependence of the phases associated with
the right and left lateral peaks, as well  as the sum of their
values. From this we can derive quantities such as the fidelity
and visibility.

The simplest phase relationship, and one that can lead
theoretically to 100\% visibility, demands that we maintain a
constant relationship, $\Phi_{jk}^R = \Phi_{jk}^L$, as we scan the
two phases. A visibility of 100\% can only be obtained for a
discrete set of integer values for $r$ which produce maximum
destructive interference, that is, such that Eq.(\ref{eq:fringes})
goes to zero. In Fig.\ref{fig:visibilities} we see the coincidence
count rate as the phases are scanned in this manner. The points
are the experimental results and the solid line is a least-squares
fit based on the function given in Eq.(\ref{eq:fringes}). We can
also use this fit to find the maximum and minimum and hence also
the visibility. Using this standard definition for the visibility
we find both the net (noise subtracted) and raw values, V$_{net}$
= 0.919 $\pm$0.026 and V$_{raw}$ = 0.815$ \pm$ 0.021.  The noise level is also shown just below the raw
interference fringe in Fig.\ref{fig:visibilities}. In the inset we
have also shown a longer section of data illustrating the control
and stability of this source.

It is clear from the form of Eq.(\ref{eq:fringes}) and
Fig.\ref{fig:visibilities} that the interference fringes resulting
from varying the two phases are qualitatively and quantitatively
different from those of qubits which depend on just one sinusoidal
function. One result of this is that we no longer have a simple
relationship between fidelity and visibility that qubits provide.
To obtain a qualitative relationship between these in the qutrit
regime we need to satisfy very specific constraints for the two
phases. If we have $r=1$ then both the visibility and the fidelity are  related and can be determined directly from the fitting parameter $\lambda$, however, the visibility is far more sensitive to the relationship between the two phases. On the other hand we can define the fidelity, with respect to a maximally entangled qutrit state, as $ F = {\rm Tr}[\rho_{jk} |\psi_{me}\rangle \langle \psi_{me}| ] = (1+8 \lambda)/9$. We can then  determine the fidelity from the value for $\lambda$,  given by the fit to  Eq.(\ref{eq:fringes}). The results for both the net and raw fidelity are F$_{net}$ = 0.984 $ \pm $ 0.018 and F$_{raw}$ = 0.843 $\pm$ 0.016. To determine the visibility in this manner we need to strictly constrain the phases. Specifically, using the definition we gave for visibility and replacing the intensities with the probabilities of Eq.(\ref{eq:fringes}), we find $V^*=3\lambda/(2+\lambda)$. Given this one might expect the visibility to be higher, $V^*= 0.982$ however we do not have the phases constrained well enough to satisfy this. By observing the left and right satellite peaks we find that the fits reveal a value of $r= 1.027 \pm 0.003$ which is close but highlights the phase sensitivity of the visibility measure over the fidelity.

\begin{figure}\leavevmode
\begin{center}
\epsfig{figure=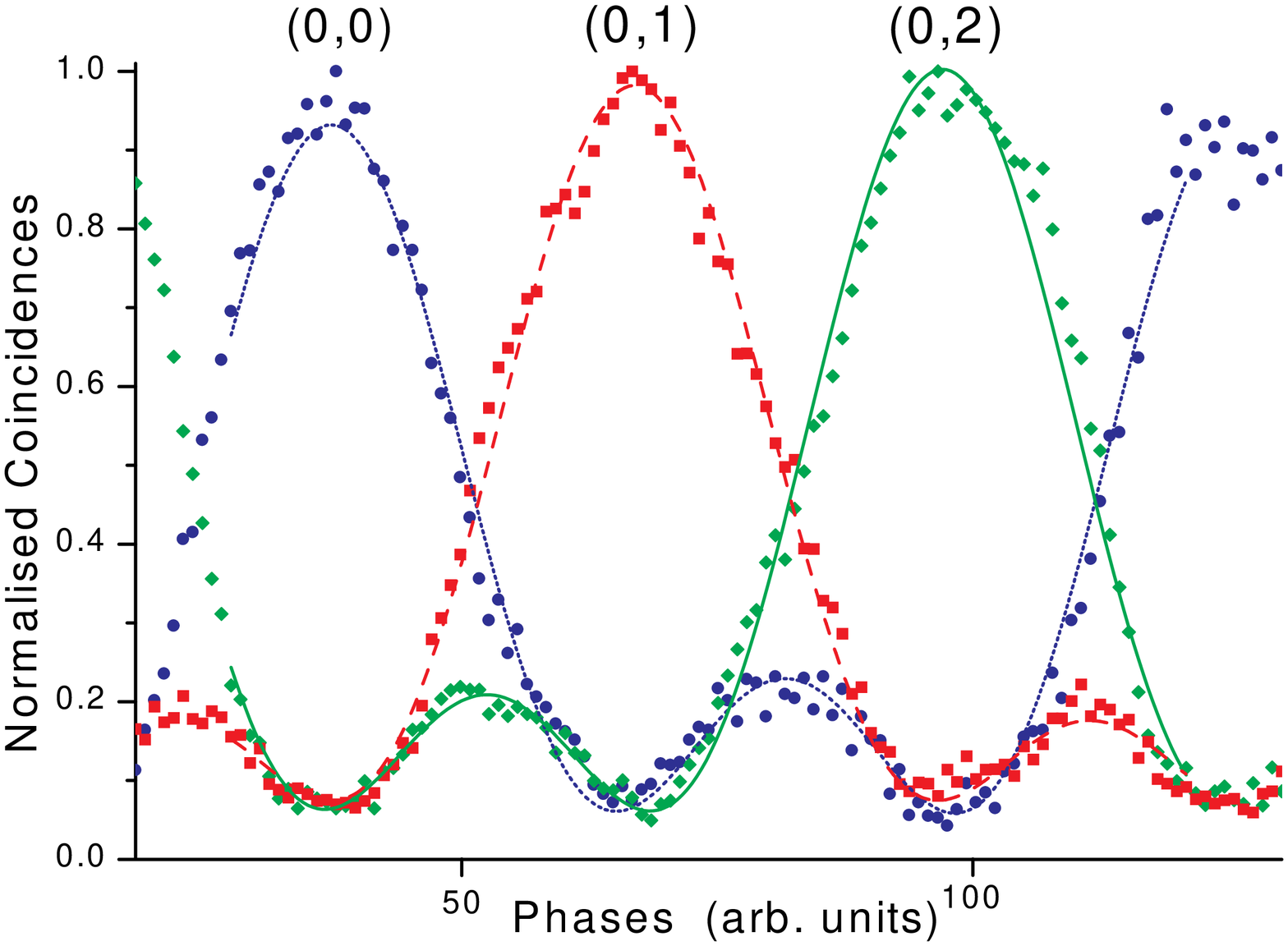,width=100mm}
\end{center}
\fcaption{\label{fig:3fringes}Normalized interference fringes, as
in Fig.\ref{fig:visibilities}, for the three possible coincidence
detections highlighting the three-fold symmetry.}
\end{figure}

In Fig.\ref{fig:3fringes} we see the normalized (accounting for
variation in detection efficiency) coincidences and fits for all
three coincidence outcomes measured at the same time. These all
have the same form as that of Fig.\ref{fig:visibilities} and are
equally distributed with respect to the phase axis. This
separation corresponds to the $t= 2\pi/3$ phase shift between the
three possible outcomes. While we clearly see the 3-fold symmetry
we also see that it is not perfect. The alignment between the
minima and maxima changes slightly from one period to the next.
This is due to the ratio between the two phases, $r$, not being
exactly one.

\section{Conclusions}

We have presented an experimental arrangement to generate, control
and measure entangled qutrits. The system is readily adaptable to
investigate both fundamental and applied aspects of high
dimensional quantum systems, for example, entanglement and
non-locality via Bell type inequalities as well as quantum key
distribution. The results show a resource capable of producing
high fidelity maximally entangled qutrits with good
controllability ideal for quantum communication.

\nonumsection{Acknowledgements}

The Authors would like to acknowledge useful discussions with
S.~P.~Kulik and his group through INTAS and the technical
assistance of J-D.Gautier. This project is financed by the Swiss
NCCR "Quantum Photonics" and the EU IST-FET project RamboQ.

\nonumsection{References}

\end{document}